\documentclass[amsmath,pra,twocolumn,showpacs]{revtex4}
\usepackage{graphics}

\begin{document}     

\title{Parametric beating of a quantum probe field with a prepared Raman coherence in a
far-off-resonance  medium}
\author{Fam Le Kien} 
\altaffiliation{On leave from Department of Physics, University of
Hanoi, Hanoi, Vietnam; also at Institute of Physics, National Center for 
Natural Sciences and Technology, Hanoi, Vietnam} 
\affiliation{Department of Applied Physics and Chemistry, 
University of Electro-Communications, Chofu, Tokyo 182-8585, Japan\\
CREST, Japan Science and Technology Corporation (JST), Chofu, Tokyo 182-8585, Japan}
\author{K. Hakuta}
\affiliation{Department of Applied Physics and Chemistry, 
University of Electro-Communications, Chofu, Tokyo 182-8585, Japan\\
CREST, Japan Science and Technology Corporation (JST), Chofu, Tokyo 182-8585, Japan}
\date{\today}

\begin{abstract}
We investigate the parametric beating of a  quantum probe field with a prepared Raman coherence
in a far-off-resonance medium, and describe the resulting  multiplexing processes.
We show that the normalized autocorrelation  functions of the  probe field 
are exactly reproduced in the  Stokes and anti-Stokes  sideband fields.  
We find that an initial coherent state of the probe field  can be replicated to the Raman sidebands, 
and an initial squeezing of the probe field can be partially transferred to the sidebands. 
We show that a necessary condition for the output fields to be in an entangled state or, more generally, in a nonclassical state is that the input field state is a nonclassical state. 
\end{abstract}

\maketitle

Recently, considerable attention has been drawn to the parametric beating of a weak probe field with a prepared Raman
coherence in a far-off-resonance medium \cite{Nazarkin99,beating,Liang,
Katsuragawa}. It has been shown that coherently excited molecular oscillations can produce 
ultrabroad Raman spectra \cite{Nazarkin99,Liang,Katsuragawa,Modulation} that may  synthesize to  subfemtosecond \cite{some theory,Sokolov01,korn02} and subcycle  \cite{HarrisSeries} pulses. 
It has been demonstrated that a multimode laser radiation  \cite{Liang} and even
an incoherent fluorescent light \cite{Katsuragawa}  can be  replicated into  Raman sidebands. 
Due to a substantial molecular coherence produced by the two-color adiabatic Raman pumping method \cite{Liang,Katsuragawa,Modulation,some theory}, the quantum conversion efficiency of the parametric beating technique can be maintained high even for weak lights with  less than one photon per wave packet \cite{Katsuragawa}. 
To describe the evolution of the statistical characteristics of such weak fields, 
quantum treatments for the fields are required. 
In related problems, the possibility of  transferring a quantum state of light with one carrier
frequency  to another carrier  frequency (multiplexing) has been discussed for resonant systems \cite{Scully},
and the generation of correlated photons using the $\chi^{(2)}$ and $\chi^{(3)}$ parametric processes has been intensively studied \cite{Mandel and Scully book,Wang}. 
However, to our knowledge,  the quantum properties of the fields in the parametric beating  with a prepared Raman coherence have not been examined.

In this paper, we investigate  the parametric beating of a  quantum probe field with a prepared
Raman coherence in a far-off-resonance medium, and describe the resulting  multiplexing processes.
We show that the normalized autocorrelation  functions of the  probe field 
are exactly reproduced in the  Stokes and anti-Stokes  sideband fields 
(autocorrelation multiplexing).  We find that
an initial  coherent state of the probe field  can be replicated to the Raman sidebands 
(coherent-state multiplexing), and an initial  squeezing of the probe field can be partially transferred to
the sidebands. 
We show that a necessary condition for the output fields to be in an entangled state or, more generally, in a nonclassical state is that the input field state is a nonclassical state. 

\begin{figure}
\begin{center}
  \includegraphics{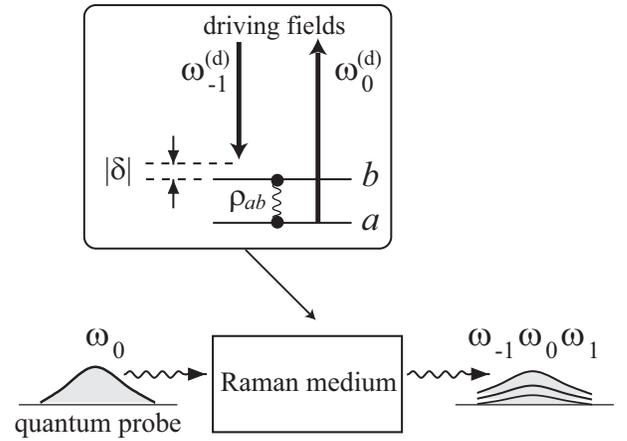}
 \end{center}
\caption{Principle of the technique: 
Two classical laser fields drive 
a Raman transition of molecules in a  far-off-resonance medium. 
The beating of a weak quantum probe field with the prepared Raman coherence produces two new sideband fields.}
\label{fig1}
\end{figure}

We consider a far-off-resonance Raman medium, see Fig. \ref{fig1}. 
We send a pair of long, strong, classical  laser fields, with carrier frequencies 
$\omega_{-1}^{(d)}$ and $\omega_0^{(d)}$,
and a short, weak, quantum probe field $\hat E_0$, with carrier frequency $\omega_0$, through the  Raman medium, along the $z$ direction. The timing and alignment of these fields are such  that  they  
substantially overlap with each other 
during the interaction process. The driving laser fields are tuned close to 
the Raman transition $|a\rangle \rightarrow |b\rangle$, with a small finite two-photon detuning $\delta$,  
but are far detuned from the upper electronic states  of the molecules. These driving fields adiabatically 
produce a Raman coherence $\rho_{ab}$ \cite{some theory}. 
When the probe field propagates through the medium, it beats with  the 
Raman coherence prepared by the  driving fields.
Since the probe field is weak, the medium state does not change substantially during this step.  
The beating of the probe field with the prepared Raman coherence results in two new quantum fields 
$\hat E_{-1}$ and $\hat E_1$, at the Stokes and anti-Stokes frequencies $\omega_{-1}$ and $\omega_1$, respectively.  
We assume that the prepared Raman coherence $\rho_{ab}$ is substantial  so that the spontaneous Raman processes
are negligible compared to the stimulated processes. 
We also assume that the product of the coherence $\rho_{ab}$ and the medium length $L$  is not too large
so that the generation of high-order sidebands of the probe field  can be neglected. 
When we take the classical propagation equations for the Raman sidebands \cite{some theory}
and replace the  field amplitudes   by the quantum operators, we obtain 
\begin{eqnarray}
\frac{\partial \hat E_0}{\partial  z }+\frac{\partial \hat E_0}{c\partial  t }
&=& i\beta_0
(a_0\hat E_0 +d_{-1}\rho_{ba}\hat E
_{-1} +d_0\rho_{ab}\hat E_1), \nonumber\\
\frac{\partial \hat E_1}{\partial  z }+\frac{\partial \hat E_1}{c\partial  t }
&=& i\beta_1
(a_1\hat E_1 +d_0\rho_{ba}\hat E _0 ) ,
\nonumber\\  
\frac{\partial \hat E_{-1}}{\partial  z }+\frac{\partial \hat E_{-1}}{c\partial  t }
&=& i\beta_{-1}
(a_{-1}\hat E_{-1} 
+d_{-1}\rho_{ab}\hat E_0). \label{1}
\end{eqnarray}
Here, $a_q$ and $d_q$ with $q=0,\pm1$ are the dispersion and coupling constants, respectively. We have denoted 
$ \beta_q= {\cal N} \hbar \omega_q/\epsilon_0 c$,  where ${\cal N}$ is the molecular number density. 

We introduce the propagation constants
$ \kappa_q=\beta_q a_q$, 
 and  define  the phase mismatch
$ \Delta k=2\kappa_0-\kappa_1-\kappa_{-1} $.
We write 
$\rho _{ab} = \rho_0 \exp[-i (\kappa_1-\kappa_{-1}) z /2 +i\phi_0]$, 
where $\rho_0\equiv|\rho_{ab}|$, and assume that $\rho_0$ and $\phi_0$ are constant in time and space.
We change the variables by 
$\hat E_0=\hat {\cal E}_0 \exp[i (\kappa_0-\Delta k/4)  z ]$ and 
$\hat E_{\pm1}=\hat {\cal E}_{\pm1} \exp[i (\kappa_{\pm1}+\Delta k/4)  z \mp i\phi_0]$.
In terms of photon operators, we write
$\hat{\cal E}_q(z,t)=(2\hbar\omega_q/\epsilon_0LA)^{1/2}\sum_K  \hat b_q(K,t) e^{iK(z-ct)}$.
Here, $L$ is the quantization length, which is taken to be equal to the medium length, 
$A$ is the quantization transverse area, which is taken to be equal to the beam area, 
$K$ is a Bloch wave vector,
and $\hat b_q$ and $\hat b_q^\dagger$ are annihilation and creation operators
for the $q$th mode. 
Then, we obtain
\begin{eqnarray}
\frac{\partial \hat b_0}{\partial  t }
&=&i\Delta \hat b_0 
+i(g_{-1} \hat b_{-1} +g_1 \hat b_1),
\nonumber\\
\frac{\partial \hat b_{\pm 1}}{\partial  t }
&=& -i\Delta  \hat b_{\pm 1} + i
g_{\pm 1}   \hat b_0 , 
\label{14}
\end{eqnarray}
where
$\Delta={c\Delta k}/{4}$,
$g_1=({\cal N} \hbar/\epsilon_0 )\sqrt{\omega_1\omega_0} \, d_0  \rho_0$, 
and $g_{-1}=({\cal N} \hbar/\epsilon_0 )\sqrt{\omega_{-1}\omega_0}\,  d_{-1}\rho_0$. 
It follows from Eqs. (\ref{14}) that the boson operator commutation relations 
$[\hat b_q,\hat b_{q'}^\dagger]=\delta_{q,q'}$ and 
$[\hat b_q,\hat b_{q'}]=[\hat b_q^\dagger,\hat b_{q'}^\dagger]=0$ are conserved in time.
We also find that
$\hat b_0^\dagger\hat b_0+\hat b_1^\dagger\hat b_1+\hat b_{-1}^\dagger\hat b_{-1}
=\textrm{const}$,
that is, the  total photon number is conserved. Note that Eqs. (\ref{14}) 
are the Heisenberg equations for the fields that are coupled to each other by the effective interaction Hamiltonian
\begin{eqnarray}
\hat H&=&\hbar \Delta (\hat b_1^\dagger\hat b_1+\hat b_{-1}^\dagger\hat b_{-1}-\hat b_0^\dagger\hat b_0) \nonumber\\&&\mbox{}
-\hbar [g_1 (\hat b_0 \hat b_1^\dagger+\hat b_1 \hat b_0^\dagger)+g_{-1} (\hat b_0 \hat b_{-1}^\dagger+\hat b_{-1} \hat b_0^\dagger)].
\quad
\label{17}
\end{eqnarray}

For simplicity, we restrict our discussion to  the case where only a single mode of the probe field (with, e.g., $K=0$) is initially excited. 
Solving Eqs. (\ref{14}), we find
\begin{equation} 
\hat b_q(t)=\sum_{q'}u_{qq'}(t)\hat b_{q'}(0),
\label{17a}
\end{equation} 
where
\begin{eqnarray}
u_{0,0}(t)&=&\cos(gt)+i(\Delta/g)\sin(gt),\nonumber\\ 
u_{1,1}(t)&=&(g_1/g_c)^2 u_{0,0}^*(t)+(g_{-1}/g_c)^2 e^{-i\Delta t}, \nonumber\\ 
u_{-1,-1}(t)&=&(g_{-1}/g_c)^2 u_{0,0}^*(t)+(g_1/g_c)^2 e^{-i\Delta t}, \nonumber\\ 
u_{-1,1}(t)&=&u_{1,-1}(t)=(g_1g_{-1}/g_c^2)\left[u_{0,0}^*(t)-e^{-i\Delta t}\right], \nonumber\\ 
u_{0,\pm 1}(t)&=&u_{\pm 1,0}(t)=i(g_{\pm 1}/g)\sin(gt).
\end{eqnarray}
Here, we have  denoted $g_c=\sqrt{g_1^2+g_{-1}^2}$ and $g=\sqrt{g_c^2+\Delta ^2}$.
When we introduce the boson operators 
$\hat b_c=(g_1\hat b_1+g_{-1}\hat b_{-1})/g_c$ and    
$\hat b_u=(g_{-1}\hat b_1-g_1\hat b_{-1})/g_c$,
we can rewrite the solution (\ref{17a}) as
\begin{equation}
\hat b_0(t)=\cos(gt)\hat b_0(0)+i\sin(gt)[(\Delta/g) \hat b_0(0)+
(g_c/g)\hat b_c(0)]
\label{28a}
\end{equation}
and
\begin{eqnarray}
\hat b_{\pm 1}(t)&=&\pm (g_{\mp 1}/g_c)e^{-i\Delta t} \hat b_u(0)
+(g_{\pm 1}/g_c)\cos(gt)\hat b_c(0)
\nonumber\\&&\mbox{}
+i\frac{g_{\pm 1}}{g}\sin(gt)[\hat b_0(0)-(\Delta/g_c)\hat b_c(0)]. 
\label{28}
\end{eqnarray}
The boson operators $\hat b_c$ and $\hat b_u$ 
describe two orthogonal modes that are mixtures of the sideband fields.
In terms of these operators, the effective  Hamiltonian (\ref{17}) has the form
$\hat H=\hbar \Delta (\hat b_u^\dagger\hat b_u+\hat b_c^\dagger\hat b_c-\hat b_0^\dagger\hat b_0) 
-\hbar g_c (\hat b_0 \hat b_c^\dagger+ \hat b_c \hat b_0^\dagger)$.
Unlike $\hat b_c$ and $\hat b_c^\dagger$,  the operators
$\hat b_u$ and $\hat b_u^\dagger$ are not coupled to the probe field 
operators $\hat b_0$ and $\hat b_0^\dagger$. 
Therefore, the modes described by $\hat b_c$ and $\hat b_u$ are called coupled and uncoupled
modes, respectively. 
The uncoupled mode  evolves  in time as a harmonic oscillator with
the frequency $\Delta$, that is, 
$\hat b_u(t)=\hat b_u(0)e^{-i\Delta  t}$.
Meanwhile, the coupled mode evolves in time as 
$\hat b_c(t)=\cos(gt)\hat b_c(0)
+i\sin(gt)[(g_c/g)\hat b_0(0)-(\Delta/g)\hat b_c(0)]$. 
In terms of the coupled and uncoupled mode operators $\hat b_c$ and $\hat b_u$, 
the expressions of the sideband operators $\hat b_1$ and $\hat b_{-1}$
are given by 
$\hat b_1=(g_1\hat b_c+g_{-1}\hat b_u)/g_c$
and $\hat b_{-1}=(g_{-1}\hat b_c-g_1\hat b_u)/g_c$.
In addition to the uncoupled mode described by $\hat b_u$, there are two other normal modes  described by
$\hat b_+ =  (\sqrt{g_-}\,\hat b_0- \sqrt{g_+}\,\hat b_c )/\sqrt{2g}$ and 
$\hat b_- =  (\sqrt{g_+}\, \hat b_0 + \sqrt{g_-}\,\hat b_c)/\sqrt{2g}$, where $g_{\pm}=g\pm \Delta $.  
They evolve in time as
$\hat b_+(t)=\hat b_+(0)e^{-igt}$
and $\hat b_-(t)=\hat b_-(0)e^{igt}$,
with the frequencies $g$ and $-g$, respectively.
The inverse transformation  yields
$\hat b_0 =  (\sqrt{g_-}\,\hat b_+ + \sqrt{g_+}\,\hat b_-)/\sqrt{2g}$ and
$\hat b_c =  (\sqrt{g_-}\,\hat b_- - \sqrt{g_+}\,\hat b_+)/\sqrt{2g}$.
In terms of the normal mode operators, the effective  Hamiltonian (\ref{17})
has the diagonal form
$\hat H=\hbar \Delta \hat b_u^\dagger\hat b_u 
+\hbar g (\hat b_+^\dagger \hat b_+ -  \hat b_-^\dagger \hat b_-)$.

We note that the interaction between  the probe field and the coupled mode field via the prepared Raman coherence  is analogous to the  interaction between the transmitted and
reflected fields from a beam splitter \cite{Mandel and Scully book}. 
However,  the two mechanisms are very different in physical nature. 
The most important
difference between them is that the two fields from the beam splitter have the same
frequency while the coupled mode in the pump-probe Raman scheme is a superposition of the two
sidebands with different Stokes and anti-Stokes frequencies.

For a Raman medium of the length $L$, the evolution time is $t=L/c$. 
The condition that  $gL/c$  is small compared to unity is required for the negligibility of the generation of  
second- and higher-order sidebands \cite{some theory}. 
In what follows we use the explicit expressions of the output operators $\hat b_{0,\pm1}(L/c)$ to calculate  various quantum statistical characteristics of the fields. 
We mostly restrict our discussion to the case where the Stokes and anti-Stokes sideband fields are initially in the vacuum state. 

First, we calculate the correlation functions of the fields. 
We assume that the probe field is initially in an arbitrary state while the two sideband fields are initially in the vacuum state.
Using Eqs. (\ref{28a}) and (\ref{28}), we can easily calculate the normally ordered photon-number moments
\begin{eqnarray}
\langle\hat b_0^{\dagger n}\hat b_0^n \rangle&=&\left[1-(g_c/g)^2\sin^2(gL/c)\right]^n\langle
\hat b_0^{\dagger n}(0)\hat b_0^n(0)\rangle,
\nonumber\\
\langle\hat b_{\pm 1}^{\dagger n}\hat b_{\pm 1}^n \rangle&=&(g_{\pm 1}/g)^{2n}\sin^{2n}(gL/c) \langle
\hat b_0^{\dagger n}(0)\hat b_0^n(0)\rangle.
\label{32c}
\end{eqnarray}
Hence, the normalized $n$th-order autocorrelation functions
$g_q^{(n)}=\langle\hat b_q^{\dagger n}\hat b_q^n \rangle /\langle\hat b_q^\dagger\hat b_q \rangle^n$ are found to be
\begin{equation}
g_1^{(n)}=g_{-1}^{(n)}=g_0^{(n)}=
\frac{\langle\hat b_0^{\dagger n}(0)\hat b_0^n(0) \rangle}{\langle\hat b_0^\dagger(0)\hat b_0(0) \rangle^n}.
\label{32b}
\end{equation}
In particular, the normalized  second-order autocorrelation functions  
are obtained as $g_1^{(2)}=g_{-1}^{(2)}=g_0^{(2)}=\langle\hat n_0^{\textrm{(in)}}(\hat n_0^{\textrm{(in)}}-1) \rangle/\langle\hat n_0^{\textrm{(in)}} \rangle^2$.
Here, $\hat n_0^{\textrm{(in)}}\equiv\hat b_0^\dagger(0)\hat b_0(0) $ is the photon-number operator for the probe field at the input.
Thus, the generated sideband fields and the probe field have  the same normalized autocorrelation functions, which are independent of the evolution time and are solely determined by the statistical properties of the input probe field. 
In other words, the normalized autocorrelation functions of the probe field do not change during the beating
process and are precisely replicated to the generated sideband fields. 
In particular, if the photon statistics of the input probe field is sub-Poissonian, Poissonian, or
super-Poissonian, the photon statistics of the sideband fields will also be sub-Poissonian, Poissonian, or
super-Poissonian, respectively.
Such a replication of the normalized autocorrelation characteristics is called  autocorrelation multiplexing.
This result is in agreement with the conclusions of the experiments on replication
of multimode laser radiation \cite{Liang} and broadband incoherent light \cite{Katsuragawa}.
The ability of the Raman medium to multiplex the  autocorrelation characteristics
is similar to but, because of the change in carrier frequency, 
somewhat different from the property of a beam splitter \cite{Mandel and Scully book}.

The normalized two-mode  cross-correlation functions are defined by
$g_{kl}^{(2)}=\langle\hat n_k\hat n_l\rangle/(\langle\hat n_k\rangle\langle\hat n_l\rangle)$,
where $k\not=l$  and  $\hat n_q=\hat b_q^\dagger\hat b_q$.  
Using Eqs. (\ref{28a}) and (\ref{28}), we find
\begin{equation}
g_{0,1}^{(2)}=g_{0,-1}^{(2)}=g_{1,-1}^{(2)}=\frac{\langle\hat n_0^{\textrm{(in)}}(\hat n_0^{\textrm{(in)}}-1) \rangle}{\langle\hat n_0^{\textrm{(in)}} \rangle^2}.
\label{35}
\end{equation}
Thus, the normalized two-mode cross-correlation functions are equal to each other and to the normalized second-order
autocorrelation functions of the fields. Note that the correlations between the modes are nonzero, that is, $g_{kl}^{(2)}\not=1$  ($k\not=l$), only if the photon statistics of the input probe field is not Poissonian. 
With respect to these properties,  the Raman scheme is also similar to  a beam splitter 
\cite{Mandel and Scully book} except for the fact that the modes in  the Raman medium 
have different frequencies.

Second, we examine  the  squeezing  of the  field quadratures in the case where the two sideband fields are initially in the vacuum state.
A field quadrature of the $q$th mode is defined by 
$\hat X_q=\hat b_q^\dagger e^{i\varphi}+\hat b_qe^{-i\varphi}$.
We say that the $q$th  mode is in a squeezed state if there exists such a phase $\varphi$ that 
$\langle(\Delta \hat X_q)^2\rangle < 1$ or, equivalently, $S_q<0$, where $S_q=\langle(\Delta \hat X_q)^2\rangle - 1 
=2[\langle \hat b_q^\dagger\hat b_q\rangle-\langle\hat b_q^\dagger\rangle\langle\hat b_q\rangle]
+[(\langle \hat b_q^2\rangle-\langle \hat b_q\rangle^2) 
e^{-2i\varphi}+\textrm{c.c.}]$.
The squeezing degree is measured by the quantity $-S_q$.
Using Eqs. (\ref{28a}) and (\ref{28}), we find
\begin{eqnarray}
S_0(\varphi)&=&[1-(g_c/g)^2\sin^2(gL/c)]S_0^{\textrm{(in)}}(\varphi-\varphi_L),
\nonumber\\
S_{\pm 1}(\varphi)&=&(g_{\pm 1}/g)^2\sin^2(gL/c) S_0^{\textrm{(in)}}(\varphi-\pi/2).
\label{38}
\end{eqnarray}
Here, $\varphi_L=\arctan[(\Delta/g)\tan (gL/c)]$ is an angle, and $S_0^{\textrm{(in)}}(\varphi)$ is the squeezing factor for the $\varphi$-quadrature of the input probe field. 
We find from Eqs. (\ref{38}) that, if $S_0^{\textrm{(in)}}(\varphi)<0$, then $S_{\pm1}(\varphi+\pi/2)<0$. Thus, if the input probe field is in a squeezed state,  then the generated sideband fields are also in squeezed states. In other words, the squeezing of the input probe
field is transfered to the sideband fields during the beating process.
The squeezing factors $S_{\pm1}(\varphi+\pi/2)$ of the sideband fields
are reduced from the squeezing factor $S_0(\varphi)$ of the input probe field by the factors $(g_{\pm1}/g)^2\sin^2(gL/c)$, and the phases of the squeezed quadratures change by $\pi/2$. This result can be used to  convert (partially)
squeezing to a new frequency, i.e., to perform partial squeezing multiplexing.
We  note that the normalized squeezing factors $s_q\equiv S_q/\langle \hat n_q\rangle$ satisfy the relations 
$s_1(\varphi+\pi/2)=s_{-1}(\varphi+\pi/2)=s_0(\varphi+\varphi_L)=s_0^{\textrm{(in)}}(\varphi)$. 
Consequently, these normalized factors have 
the same maximal and minimal values for all the three fields.  
We also note that, when the input probe field is in a coherent state, the sideband fields have no squeezing. This property is similar to the case of four-wave mixing but is unlike
the case of degenerate parametric down-conversion, where perfect squeezing can  in principle be obtained. 
  
Next, we calculate the state of the fields in 
a special case where the input probe field is in a coherent state $|\alpha\rangle_0$
and the two sideband fields are initially in the vacuum state.
The  state of the fields at the input is 
\begin{equation}
|\Psi_{\textrm{in}}\rangle=|0\rangle_{-1}|\alpha\rangle_0|0\rangle_1 =e^{-|\alpha|^2/2}e^{\alpha \hat b_0^\dagger(0)}|0\rangle.
\label{39}
\end{equation}
The state of the fields at the output is given by 
$|\Psi_{\textrm{out}}\rangle=\hat U(L/c) |\Psi_{\textrm{in}}\rangle$,
where $\hat U(L/c)=\exp(-i\hat HL/\hbar c)$.
Since $\hat U(L/c)|0\rangle=|0\rangle$, we have
$|\Psi_{\textrm{out}}\rangle=e^{-|\alpha|^2/2}e^{\alpha \hat b_0^\dagger(-L/c)}|0\rangle$.
Using Eq. (\ref{28a}), we find 
\begin{equation}
|\Psi_{\textrm{out}}\rangle=|\alpha_{-1}(L/c)\rangle_{-1}|\alpha_0(L/c)\rangle_0 |\alpha_1(L/c)\rangle_1 . 
\label{45}
\end{equation}
Here, $|\alpha_q(L/c)\rangle_q$ is a coherent state of the $q$th mode, with the amplitude   
$\alpha_q(L/c)=\alpha u_{0q}(L/c)$. 
Thus, a probe field in a  coherent state 
can produce two sideband fields that are also in coherent states. 
Such a process is called coherent-state multiplexing.
This  property of the Raman medium is also similar to that of a beam splitter \cite{Mandel and Scully book}
except for the fact that there is a change of the carrier frequency in the Raman scattering process.

Finally, we calculate the state of the output field for the case where the input state of the probe field is a Fock state $|n\rangle_0$ and the input state of the  sideband fields is the vacuum state.
The  state of the  fields at the input is 
\begin{equation}
|\Psi_{\textrm{in}}\rangle=|0\rangle_{-1}|n\rangle_0 |0\rangle_1=\frac{\hat b_0^{\dagger n}(0)}{\sqrt{n!}} |0\rangle.
\label{46}
\end{equation}
The  state of the fields at the output is given by    
$|\Psi_{\textrm{out}}\rangle=[\hat b_0^{\dagger n}(-L/c)/\sqrt{n!}] |0\rangle$.
With the help of Eq. (\ref{28a}),  the state of the fields at the output is found to be
\begin{eqnarray}
|\Psi_{\textrm{out}}\rangle&=& 
\sum_{l=0}^n \sum_{m=0}^l\left(\begin{array}{c}n\\ l\end{array}\right)^{1/2}
\left(\begin{array}{c}l\\ m\end{array}\right)^{1/2}\nonumber\\&&\mbox{}\times
u_{0,-1}^{l-m}(L/c) u_{0,0}^{n-l}(L/c) u_{0,1}^m (L/c)
\nonumber\\&&\mbox{}\times
|l-m\rangle_{-1}|n-l\rangle_0 |m\rangle_1 .
\label{48}
\end{eqnarray}
In general, expression (\ref{48}) stands for a tripartite entangled state. 
Tripartite entangled states can find various applications in quantum information \cite{Zeilinger,Chuan}. 
 
More generally, we can show that an arbitrary coherent state $|\alpha_{-1}(0),\alpha_0(0),\alpha_1(0)\rangle$ of the input fields produces a  coherent state $|\alpha_{-1}(L/c),\alpha_0(L/c),\alpha_1(L/c)\rangle$ of the output fields,
where $\alpha_q(L/c)=\sum_{q'}u_{qq'}(L/c)\alpha_{q'}(0)$. 
Consequently, the diagonal coherent-state representation $P_{\textrm{in}}(\{\alpha_q\})$ of an arbitrary state
$\hat\rho_{\textrm{in}}$ of the input fields  determines  the  representation $P_{\textrm{out}}(\{\alpha_q\})$ 
of the state $\hat\rho_{\textrm{out}}$ the output fields   via the equation 
\begin{equation}
P_{\textrm{out}}(\{\alpha_q\})=P_{\textrm{in}}(\{\sum_{q'}u_{qq'}^*(L/c)\alpha_{q'}\}).
\label{49}
\end{equation}
If the input state $\hat\rho_{\textrm{in}}$ is a classical state \cite{Mandel and Scully book}, $P_{\textrm{in}}(\{\alpha_q\})$ must be 
non-negative  and less singular than a delta function,  and 
consequently so must $P_{\textrm{out}}(\{\alpha_q\})$.
In this case, the output state $\hat\rho_{\textrm{out}}$ is also a classical state. Moreover,  
since the coherent state $|\alpha_{-1},\alpha_0,\alpha_{-1}\rangle$ is separable and the weight factor $P_{\textrm{out}}(\{\alpha_q\})$ is non-negative, the output
state $\hat\rho_{\textrm{out}}$ is,  by definition, separable \cite{Chuan,Zeilinger}. 
Therefore, a necessary condition for the output fields
to be in an entangled state or, more generally, in a nonclassical state 
is that the input field state is  a nonclassical state. 
A similar condition has been derived for the beam splitter entangler \cite{Knight}. 
Note that, in the case where the two sideband fields are initially in the vacuum state,
Eq. (\ref{49}) reduces to
$P_{\textrm{out}}(\{\alpha_q\})=
P_{\textrm{in}}^{(0)}(\sum_{q}u_{0q}^*\alpha_{q})
\delta(\sum_{q}u_{1q}^*\alpha_{q})
\delta(\sum_{q}u_{-1q}^*\alpha_{q})$, where $P_{\textrm{in}}^{(0)}(\alpha)$ is the
coherent-state representation of an arbitrary state of the input probe field.

In conclusions, we have shown  that the parametric beating of a quantum probe field with
a prepared Raman coherence can replicate the normalized autocorrelation functions into the
Stokes and anti-Stokes sidebands.   
We have found that
an initial  coherent state of the probe field  can be replicated to the Raman sidebands, 
and an initial  squeezing of the probe field can be partially transferred to  
the sidebands. 
We have shown that a necessary condition for the fields
at the output of the Raman medium to be in an entangled state or, more generally, in a 
nonclassical state is that the input field state is a nonclassical state. 
We emphasize that, for the validity of our model, 
the prepared Raman coherence should  be substantial (so that the effect of the probe on the coherence is negligible),
but the product of the coherence  and the medium length 
should  not be too large (because otherwise high-order sidebands would be involved).   
Such Raman excitation level   can be 
produced by applying the two-color adiabatic Raman pumping technique 
for far-off-resonance Raman media, such as solid hydrogen,  molecular hydrogen gas, and deuterium gas \cite{Liang,Katsuragawa,Modulation,some theory}.
The  pump-probe Raman scheme is characterized by high conversion efficiency and
therefore can provide  a good tool to replicate quantum statistical
characteristics from one mode to another.

Useful discussions with S. Dutta-Gupta and A. K. Patnaik are  gratefully acknowledged.

\end{document}